\documentclass[twocolumn,aps,prb,ltxgrid]{revtex4}

\usepackage{epsfig}
\usepackage{graphicx}
\usepackage{amsbsy,amssymb,amsmath,bm,ulem}
\usepackage[usenames]{color}

\newcommand{\greekbf}[1]{\ensuremath{\mbox{\boldmath{$#1$}}}}

\newcommand{\f}[1]{Fig.~\ref{#1}}

\newcommand{\eqs}[2]{Eqs.~(\ref{#1}) and~(\ref{#2})}

\begin{document}

\title{Interaction between superconducting vortices and a Bloch wall
  in ferrite garnet films} 

\author{J. I. Vestg\aa rden, D. V. Shantsev, \AA. A. F. Olsen,
Y. M. Galperin, V. V. Yurchenko, P. E. Goa and T. H. Johansen}
\affiliation{Department of Physics and Center for Materials
Science and Nanotechnology,    University of Oslo, P. O. Box 1048
Blindern, 0316 Oslo, Norway}

\begin{abstract}
A theoretical model for how Bloch walls occurring in
in-plane magnetized ferrite garnet films can serve as efficient
magnetic micro-manipulators is presented. 
As example, the walls'
interaction with Abrikosov vortices in a superconductor in close contact
with a garnet film is analyzed within the London
approximation.
The model explains how vortices are attracted to
such walls, and excellent quantitative agreement is obtained for
the resulting peaked flux profile determined experimentally in
NbSe$_2$ using high-resolution magneto-optical imaging of vortices.
In particular, this model, when generalized to include charged
magnetic walls, explains the counter-intuitive attraction observed
between vortices and a Bloch wall of opposite polarity.
\end{abstract}

\maketitle

Microscopic magnetic potentials offer an efficient and often
indispensable way to manipulate various tiny objects, e.g., to trap
and guide Bose-Einstein condensates and degenerate Fermi gases of
ultracold atoms~\cite{hansel01,ott01,aubin06} or to stretch and
twist DNA strands attached to paramagnetic
beads~\cite{strick96,gosse}. How precisely one can position the
trapped object and how large forces can be applied are determined
by the steepness of the magnetic potential produced by the
manipulating device. Typically, an assembly of microfabricated
wires or coils can generate fields with gradients up to
$10^2-10^3$~T/m~\cite{hansel01,ott01,folman00,haber00}. Recently,
magnetic manipulators were created using ferrimagnetic films,
where domain walls create locally even stronger field gradients.
 In particular, ferrite-garnet films (FGFs) were used to
trap ultracold neutral atoms~\cite{shevchenko}, assemble and guide
colloidal particles~\cite{hel05pyr,hel03tip}, and manipulate
vortices in superconductors~\cite{goa03}. Using in-plane magnetized
FGFs, where domains are separated by Bloch walls, has two strong
advantages: (i) it is easy to move the domain wall and thereby
tune the magnetic potential, and (ii) one can simultaneously
observe the motion of the manipulated objects. This ``{\em see
what you do}'' ability stems from the large Faraday effect in the
FGFs, which today are extensively used as magneto-optical imaging
(MOI) sensors~\cite{jooss}. In an optical polarizing microscope
configuration the FGF allows direct visualization of both the
stray field from the manipulated magnetic objects and the Faraday
rotation in the wall itself.

In this work we present a theoretical model for how Bloch walls
can function as magnetic manipulators, using as example their
interaction with a lattice of Abrikosov vortices in a type-II
superconductor. The configuration considered is that of a FGF
located near, but a finite distance from the surface of a flat
superconductor.  It is shown that the model, generalized to
include charged walls, explains how the vortices are attracted to
such walls, and predicts quantitatively the non-uniform flux
density distribution we find experimentally using MOI.
\begin{figure}[t]
  \epsfig{file=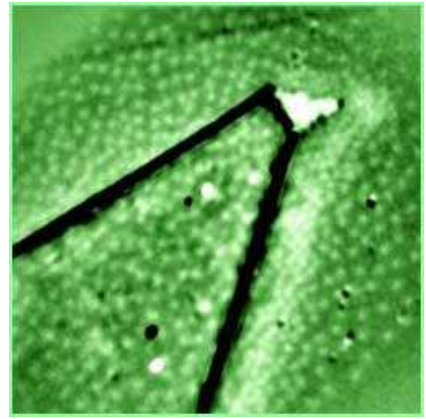, width=5.5cm}
  \caption{Magneto-optical image of the vortex distribution near a
    Bloch wall in a Bi-substituted lutetium iron garnet film placed on top of a
    superconducting NbSe$_2$ crystal with transition temperature of 7.2~K.
    The slightly uncrossed analyzer and polarizer setting used
    here implies that the dark wall and the bright
    vortices have opposite field polarities.
    Image dimensions are 70$\times$70~$\mu$m$^2$.}
  \label{fig:MO}
\end{figure}

\f{fig:MO} shows an image of the magnetic field distribution near
two linear Bloch wall segments, which themselves appear dark in
the image. The superconductor is a single crystal of 
NbSe$_2$\cite{oglesby94}
cooled to $T = 4~$K in a 0.3~mT perpendicular magnetic field,
which transformed into quantized vortices when entering the
superconducting state. Each vortex is here seen as a bright dot.
Evidently, the magnetic wall has a considerable attraction on the
vortices, since their number density increases near the wall. This
observation forms the experimental basis for the theoretical
modeling presented below. Interestingly, a most surprising
feature visible from \f{fig:MO} makes the problem even more
challenging. There is opposite contrast between the dark wall and
the bright vortices, which means they are of opposite magnetic
polarity. In this case one would expect from the models presented
previously in the literature~\cite{helseth02,milosevic02,burmistrov05, helseth02-2}  that a Bloch wall should {\em repel}
the vortices. As will be shown, by accounting for additional
magnetic charges due to misalignment between the wall direction
and the in-plane magnetization vector within the domains, the sign
of the interaction can become inverted.

The two Bloch wall segments seen in \f{fig:MO}
are actually part of a larger zigzag pattern. Extended zigzag domain walls
are commonly present in FGFs with strong in-plane anisotropy~\cite{VV76}.
An example is shown in \f{fig:sample}~(top), where the zigzag line separates
two domains with antiparallel magnetizations that meet head-on. By
folding into a zigzag pattern the domain boundary reduces the
density of magnetic charges at the wall, which helps to minimize
the energy~\cite{magneticdomains}.
To describe the interaction between one segment of such a zigzag
wall and a superconducting vortex, we introduce the model
illustrated in \f{fig:sample}~(bottom).
\begin{figure}[t]
  \centering
  \epsfig{file=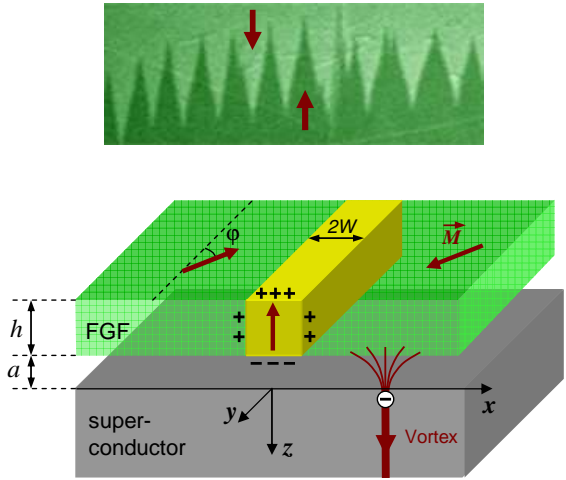, width=8.0cm}
  \caption{Top: MO image showing a zigzag Bloch wall in a FGF
    separating two domains with  antiparallel in-plane
    magnetization. Bottom: Sketch of an in-plane magnetized FGF
    with a Bloch wall placed above a superconductor. The magnetic charges
    along the vertical sides of the wall can lead to a net attraction
    between a wall and vortices, as seen in \f{fig:MO}.}
  \label{fig:sample}
\end{figure}
The superconductor occupies the half-space, $z>0$, and  the wall
is directed along the $y$ axis, which forms an angle $\varphi$
with the magnetization direction of the two domains. 
Experimentally we detect only tiny changes of the wall width, $2W$, 
during cooling through $T_c$, in agreement with~\onlinecite{helseth02}.
Thus, in calculations the wall is approximated as a fixed,
uniform, out-of-plane magnetization $M_z(|x|<W)= - M_s$. 
Inside the domains there is an
in-plane magnetization with a component normal to the wall given
by $M_x(|x|>W)=-M_s(x/|x|) \sin \varphi $, where $M_s$ is the
saturation magnetization. For $ \varphi=0^\circ$ the present model
reduces to the non-charged wall case. The $M_y$ component is
omitted in the analysis since the wall is assumed to be infinitely
long.

Stray fields from the wall induce shielding currents in the
superconductor, which we determine using the London theory. The
equations valid inside and outside the superconductor then read
\begin{equation}
  \label{sc+bw}
  \begin{array}{rll}
    \lambda^{-2}A - \nabla^2 A &=0 &,~z\geq 0 \\
    -\nabla^2 A &=\mu_0(\nabla\times\mathbf M)_y &,~z\leq 0
  \end{array}
\end{equation}
where $\lambda$ is the London penetration depth and the vector potential is
$\mathbf A = A\hat y$.
The shielding currents flow in the $y$ direction and their density
equals $J_y=-A/\mu_0\lambda^2$.
A vortex present in the superconductor then feels two forces.
First, the direct force from the FGF, which can be calculated from
the free energy term, $\mu_0\int\mathbf M\cdot \mathbf H_v dV$,
where $H_v$ is the stray field from the vortex. Second, the
Lorentz force from the shielding currents in the superconductor,
$\mathbf F_L=\mathbf J(\mathbf r)\times \greekbf \Phi_0$
integrated over the length of the vortex. $\Phi_0$ is the magnetic
flux quantum, and we will simplify the treatment by assuming the
vortex to be straight and aligned with the $z$ axis.
Interestingly, the two forces turn out to have exactly the same
magnitude and direction, as was noted also in
Ref.~\onlinecite{milosevic02} where a similar configuration was
analyzed.
\begin{figure}[t]
  \centering
 \epsfig{file=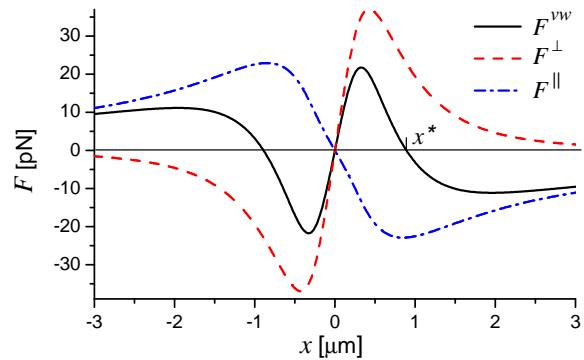, width=7.8cm}
  \caption{The calculated forces on the vortex from the Bloch
    wall: $F^{\bot}$ is repulsive and $F^{\|}$ is attractive.
    Their sum $F^{vw}$ changes sign at
    $x^* \approx 1~\mu$m.
    The parameters used here are:
    $\sin \varphi=0.34$, $2W=0.6~\mu$m, $h=0.8~\mu$m, $a=140~$nm,
    $\lambda=200~$nm, and $M_s=50~$kA/m.}
  \label{fig:F}
\end{figure}

It is convenient to express the total force on the vortex in the $x$
direction as 
\begin{equation}
  F^{vw} =  F^{\bot} +  F^{\|}
\end{equation}
where $F^{\bot}$ and $F^{\|}$ are the contributions from the
perpendicular and in-plane components of $\mathbf M$, respectively.
Their Fourier transforms are obtained as
\begin{align}
  \label{force1}
   F^{\bot}_k &=  - 4i~\frac{M_s\Phi_0}{\lambda^2}\,
   \frac{1-e^{-|k|h}}{|k|\tau (\tau+|k|)}~e^{-|k|a}~\sin Wk \, , \\
   \label{force2}
   F^{\|}_k &= 4i~\frac{M_s\Phi_0}{\lambda^2}\,
   \frac{1-e^{-|k|h}}{k\tau(\tau+|k|)}~e^{-|k|a}~\cos Wk~\sin \varphi
\end{align}
where $\tau=\sqrt{\lambda^{-2}+k^2}$,
$a$ is the gap between superconductor and FGF, and $h$ is the FGF thickness.
For the configuration illustrated in \f{fig:sample}
 the force $F^{\|}$ is always attractive, whereas $F^{\bot}$ is repulsive.
This qualitative result can be easily understood by considering
the interaction between the magnetic charges involved. The stray
field from a vortex is closely approximated by that of a magnetic
monopole located at $z \sim \lambda$  and with strength
$-2\Phi_0$~\cite{Carneiro}. Thus, the vortex is attracted to the
positive charges along the vertical sides of the wall and repelled
by the perpendicular dipole charges. The charge representation
also yields the correct magnitude of forces given by
\eqs{force1}{force2} in the limit $\lambda \to 0$. The
superconductor then perfectly screens the magnetic field created
by $\mathbf M$ and its presence should be accounted for by adding
the corresponding mirror charges.

From the inverse transform of \eqref{force1} and \eqref{force2},
we obtain the spatial variation of the two force contributions,
which are plotted in \f{fig:F} together with the total force on
the vortex. At sufficiently large $x$ the magnitude of $F^{\|}$ is
larger than that of $F^{\bot}$, which is expected since the
monopole-monopole interaction should dominate at long distances.
However, at short distances $F^{\bot}$ becomes dominant, and the
total force changes sign at some distance $x^*$. For our set of
parameters, the repulsive region $|x|<x^*$ is very small, less
than $1~\mu$m. This is why the repulsive region under the wall is
not visible in the image of \f{fig:MO}. This also explains why we
observe the counter-intuitive attraction between the Bloch wall
and the vortices of opposite polarity.
A related phenomenon when ferromagnetic domain wall 
stimulates
superconductivity due to its stray fields was considered 
in Ref.~\onlinecite{yang04}.
\begin{figure}[t]
  \centering
  \epsfig{file=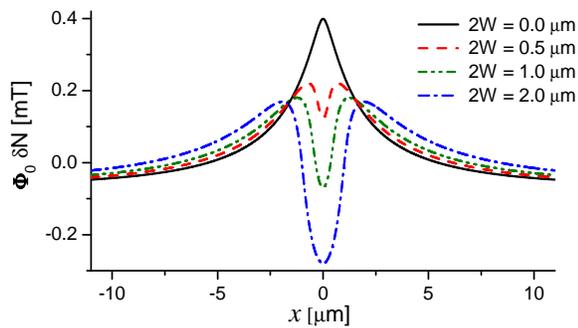, width = 7.8cm}
  \caption{The excess vortex density $\delta N(x)$ near the Bloch wall for
    various wall width $2W$ calculated using $L=160~\mu$m and the other
    parameter values as indicated in the caption of \f{fig:F}.
  }
  \label{fig:N2}
\end{figure}

We consider next an initial state with a uniform distribution of
vortices in the superconductor, and a subsequent introduction of a
Bloch wall. This results in a perturbation of the vortex
density, $\delta N(x)$, which creates an additional force acting
on every vortex. In the equilibrium, the additional force
everywhere balances the force from the Bloch wall, i.e.,
\begin{equation}
  \label{shielding}
  F^{vw}(x) = \int  F^{vv}(x-x',y')\, \delta N(x')\,  dx' dy' \ ,
\end{equation}
where $F^{vv}$ is the $x$ component of the repulsive vortex-vortex
interaction, and the vortex matter is considered as a continuum.
This equation represents a perfect {shielding}  of the domain wall
by the vortex matter. In Fourier space, the perturbed vortex
density becomes
$  \delta N_k=F^{vw}_k / F^{vv}_{k}$ .
The vortex-vortex interaction
can be obtained from the currents around a flux line
in a half-space, first calculated by Pearl \cite{pearl66},
and gives
\begin{equation}
  \label{vortexvortexforce}
  F^{vv}_{k}
  = \frac{\Phi_0^2}{\mu_0~\lambda^2 }\frac{~ik}{|k|\tau^2}
    \left(
      |k|L+\frac{1}{\lambda^2}\frac{1}{\tau}\frac{1}{|k|+\tau}
    \right)
\end{equation}
where $L$ is the flux line length. The term proportional to $L$ is
the conventional (Abrikosov) bulk contribution, while the other
term is the surface contribution.

The resulting perturbation $\delta N(x)$ of the vortex density
induced by a Bloch wall is shown in \f{fig:N2}. Its profile is
strongly dependent on the wall width~$2W$. In the small $W$ limit
the total force on a vortex is everywhere attractive, hence the
vortex density
 increases monotonously as one approaches the wall. For increasing
$W$ the density profile develops a minimum below the center of the
wall, and the resulting $\delta N(x)$ becomes non-monotonous with
a minimum at the center and maxima near the two wall edges. For
sufficiently large $W$ the vortices right below the wall are
expelled creating a narrow depleted area.
\begin{figure}[t]
  \centering
  \epsfig{file=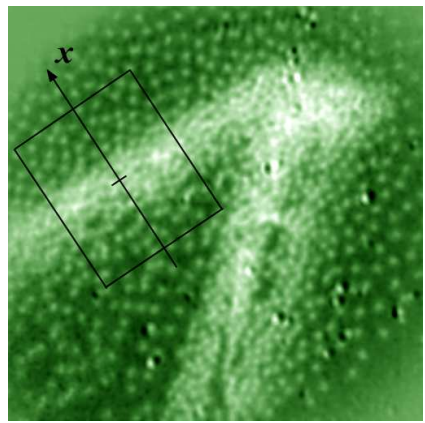, width=5.5cm} \\[3mm]
  \epsfig{file=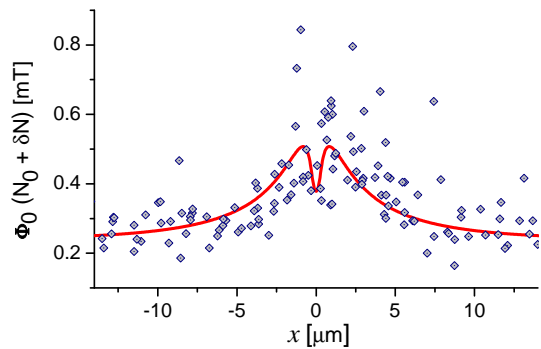, width = 7.5cm}
  \caption{Top: Distribution of vortices formed in the presence of a
    Bloch wall. The image was taken after the wall, seen in
  \protect{\f{fig:MO}}, was removed by an in-plane field of the order of a few
  $\mu$T applied perpendicular to the indicated x-axis.
    Bottom: Vortex density obtained from the image (each symbol
  represents one vortex) 
    together with the theoretical curve calculated for
    $L=160~\mu$m, $\Phi_0N_0=0.3~$mT and remaining
    parameters as listed in the caption of \f{fig:F}.}
  \label{fig:N}
\end{figure}

The theoretical vortex density profile $\delta N(x)$ can be
compared to our MOI observations of the vortex distribution near
the Bloch wall. The FGF had a thickness $h=0.8~\mu$m, and
saturation magnetization $M_s=50$~kA/m. The film, with no
additional layers (contrary to standard MOI indicators), was
placed directly on top of the 0.3~mm thick NbSe$_2$ crystal with a
gap of $a=140$~nm, as determined from the optical interference
pattern~\cite{goa03-2}.
This gap equals one quarter of a wavelength,and gives
optimal transmission.

The vortices were formed slightly below $T_c$ where flux pinning
is negligible. Thus, one expects that the vortex positions seen in
\f{fig:MO} represent a frozen picture of an arrangement where the
vortices adjust only to balance the interaction with the wall. To
obtain a better view, we made use of the fact that further cooling
to 4~K increased the vortex pinning considerably, and removed the
Bloch wall without creating noticeable change in the vortex
positions, see and \f{fig:N} (top). From this image, the positions
of all the vortices inside the marked rectangular area were
identified, and the Wigner-Seitz cell of each vortex was
determined using standard triangulation~\cite{barber96}. The local
vortex density was obtained by inverting the cell area, and is
shown in \f{fig:N}~(bottom) for every vortex versus its coordinate
$x$. The vortex density near the wall clearly exceeds the background
density of $\Phi_0 N_0 \approx 0.3$~mT. The theoretical curve
$N_0+\delta N(x)$, plotted in \f{fig:N} (bottom), was calculated
using $\varphi=20^\circ$ determined from \f{fig:MO}, and a wall width
of $2W=0.6~\mu$m.
The calculated curve
reproduces very well not only the sign, but also the magnitude of
the excess vortex density. This agreement was achieved using 
the penetration depth $\lambda=200$~nm as an adjustable
parameter. This value of $\lambda$
corresponds to the temperature of 6.8~K (slightly below $T_c=7.2$~K) which is 
thus the temperature when the vortices got frozen. 
The vortex length $L$=160~$\mu$m was another free parameter. 
It is smaller than the crystal thickness 300~$\mu$m to compensate for
the overestimation  
of the Abrikosov interaction term in the continuum approximation.

An open question remains regarding the large apparent width of
the Bloch wall, approximately 3~$\mu$m as seen from the image
in~\f{fig:MO}.
The theoretical
estimate obtained by minimizing the sum of
exchange, anisotropy and magnetostatic energies
is given implicitly by the equation\cite{helseth02}
$(1+w)^{-2}+\alpha w^{-2}=1-K_u/\mu_0 M_s^2$, where
$w=2W/h$ is the normalized wall width and $\alpha=2 \pi^2 A/\mu_0 M_s^2 h^2$.
Substituting the effective exchange constant $A=2\times
10^{-11}~$J/m and 
the uniaxial anisotropy constant $K_u\sim 10^3$~J$/$m$^3$ 
[\onlinecite{helseth02}] we obtain
 $2W\approx 0.6~\mu$m. 
The discrepancy 
between the observed and estimated wall width
is probably due to the optical diffraction which 
significantly distorts the image of objects of the order of the light wavelength, 0.55~$\mu$m. 

In conclusion, mobile domain walls found in in-plane magnetized
ferrite garnet films were investigated for possible use as
magnetic micro-manipulators. It was shown, choosing
superconducting vortices as a case example, that such films can
serve to both apply forces and simultaneously monitor the results
of the action. A theoretical model for the interaction was
developed, with the vortices described within the London
approximation, and the domain wall represented by a charged
magnetic wall. The charged wall model, which includes magnetic
poles on all the sides of the wall's rectangular cross-section, is
shown to give a very good quantitative description of the
attraction of vortices to such a wall. The comparison was made by
direct observation of individual vortices using the
magneto-optical imaging technique.

This work was supported financially by The Norwegian Research
Council, Grant No. 158518/431 (NANOMAT) and by FUNMAT@UIO. We
gratefully acknowledge discussions with 
V. Vlasko-Vlasov, L. Uspenskaya, and E. Il'yashenko.

\bibliographystyle{revtex}

\end{document}